\documentclass{aa}
\usepackage{graphicx}
\usepackage[varg]{txfonts}
\usepackage{cases}
\usepackage{natbib}

\begin{document}
   \title{Longitudinal filament oscillations enhanced by two C-class flares}

   \author{Q. M. Zhang\inst{1,2,3}, J. H. Guo\inst{3,4}, K. V. Tam\inst{2}, and A. A. Xu\inst{2}}

   \institute{Key Laboratory of Dark Matter and Space Science, Purple Mountain Observatory, CAS, Nanjing 210033, China \\
              \email{zhangqm@pmo.ac.cn}
              \and
              State Key Laboratory of Lunar and Planetary Sciences, Macau University of Science and Technology, Macau, China \\
              \and
              School of Astronomy and Space Science, Nanjing University, Nanjing 210023, China \\
              \and
              Key Laboratory of Modern Astronomy and Astrophysics (Nanjing University), Ministry of Education, Nanjing 210093, China \\
              }

   \date{Received; accepted}
    \titlerunning{Longitudinal oscillations of an active region filament}
    \authorrunning{Zhang et al.}

 \abstract
   {Large-amplitude, longitudinal filament oscillations triggered by solar flares have been well established. However, filament oscillations enhanced by flares have never been reported.}
   {In this paper, we report the multiwavelength observations of a very long filament in active region (AR) 11112 on 2010 October 18. The filament was composed of two parts,
   the eastern part (EP) and western part (WP). We focus on longitudinal oscillations of the EP, which were enhanced by two homologous C-class flares in the same AR.}
   {The filament was observed in H$\alpha$ wavelength by the GONG and in 
   extreme ultra-violet (EUV) wavelengths by the Atmospheric Imaging Assembly (AIA) on board the Solar Dynamics Observatory (SDO).
   Line-of-sight magnetograms were provided by the Helioseismic and Magnetic Imager (HMI) on board SDO.
   The global three-dimensional (3D) magnetic fields were obtained using the potential field source surface modeling.
   Soft X-ray light curves of the two flares were recorded by the GOES spacecraft. White-light images of the corona were observed by the LASCO/C2 coronagraph on board SOHO.
   To reproduce part of the observations, we perform one-dimensional, hydrodynamic numerical simulations using the MPI-AMRVAC code.}
   {The C1.3 flare was confined without a CME. Both EP and WP of the filament were slightly disturbed and survived the flare. After 5 hrs, eruption of the WP generated a C2.6 flare and a narrow, jet-like CME.
   Three oscillating threads (thd$_a$, thd$_b$, thd$_c$) are obviously identified in the EP and their oscillations are naturally divided into three phases by the two flares.
   The initial amplitude ranges from 1.6 to 30 Mm with a mean value of $\sim$14 Mm.
   The period ranges from 34 to 73 minutes with a mean value of $\sim$53 minutes. The curvature radii of the magnetic dips are estimated to be 29 to 133 Mm with a mean value of $\sim$74 Mm.
   The damping times ranges from $\sim$62 to $\sim$96 minutes with a mean value of $\sim$82 minutes. The value of $\tau/P$ is between 1.2 and 1.8.
   For thd$_a$ in the EP, the amplitudes were enhanced by the two flares from 6.1 Mm to 6.8 Mm after the C1.3 flare and further to 21.4 Mm after the C2.6 flare.
   The period variation as a result of perturbation from the flares was within 20\%. The attenuation became faster after the C2.6 flare.}
   {To the best of our knowledge, this is the first report of large-amplitude, longitudinal filament oscillations enhanced by flares. 
   Numerical simulations reproduce the oscillations of thd$_a$ very well. The simulated amplitudes and periods are close to the observed values, 
   while the damping time in the last phase is longer, implying additional mechanisms should be taken into account apart from radiative loss.}


 \maketitle

\section{Introduction}
Solar prominences are dense and cool plasma structures suspending in the corona \citep[][and references therein]{lab10,mac10,par14,vial15,gib18}. 
The density of prominences is two orders of magnitude larger than the corona, 
while the temperature is two orders of magnitude lower than the corona. Prominences are also called filaments that appear darker than the surroundings on the solar disk \citep{eng76,mar98}. 
Prominences (or filaments) can be observed in H$\alpha$, Ca\,{\sc ii}\,H, He\,{\sc i}\,10830 {\AA}, extreme-ultraviolet (EUV), and radio wavelengths \citep[e.g.,][]{van04,ber10,sch10,sch14,shen15,yan15,yang17}. 
It is generally accepted that the gravity of prominences is balanced by the upward tension force of magnetic dips 
in sheared arcades \citep{xia12,kep14,gu15}, magnetic flux ropes \citep{dev00,mar01,su12}, or both \citep{liu12}.
Filaments are located in filament channels in active regions (ARs), quiet region, and polar regions. The origins of material in filaments include direct injection from the chromosphere \citep{li13}, 
upward elevation from below \citep{li05}, evaporation-condensation \citep{xia11}, and reconnection-condensation \citep{li18b}. The lifetimes of filaments range from a few hours to several days. 
After destabilization, they are likely to erupt and generate flares and/or coronal mass ejections (CMEs) \citep{shi95}. 
When the compression from the large-scale, overlaying magnetic field lines is strong enough, the filament fails to erupt successfully and evolves into a CME \citep{ji03,zqm15}.

After being disturbed, filaments are prone to oscillate periodically \citep{oli02,arr11,arr12}.
The period of oscillations ranges from a few minutes to tens of minutes or even hours \citep{jing06,lin07}.
In most cases, they oscillate with small amplitudes and short periods of 3$-$5 minutes \citep{ning09,li18a}.
Large-amplitude oscillations are often observed as a result of sudden attacks, 
such as flares \citep{ram66}, microflares \citep{jing03,vrs07}, coronal jets \citep{luna14,zqm17b}, surges \citep{chen08}, shock waves \citep{shen14b,pant16}, 
and global EUV waves associated with CMEs \citep{eto02,shen14a}.

Due to various kinds of disturbances, the direction of filament oscillations may change from case to case \citep{luna18}. For longitudinal oscillations, the filament
material oscillates along the threads with small angles of 10$^{\circ}$-20$^{\circ}$ between the threads and spine \citep{jing03,vrs07,zqm12,li12,bi14,chen14,luna14,luna17,zhou18,awa19,zap19}.
The polarization of transverse oscillations could be horizontal (vertical) if the whole body oscillates parallel (vertical) to the solar surface \citep{hyd66,kle69,zqm18}.
Sometimes, two filaments experience different types of oscillations due to different angles between the incoming waves and filaments \citep{shen14b}. 
Occasionally, different parts of a whole filament experience different types of oscillations \citep{pant16,zqm17b,maz20}. 
Interestingly, the parameters of oscillations, including amplitude, period, and damping time vary with time, possibly due to the thread-thread interaction \citep{zqm17a,zhou17}.
Hence, filament oscillations show complex and complicated behaviors.
 
The primary restoring force of large-amplitude longitudinal filament oscillations is believed to be the gravity of filament \citep{luna12,lunas12,zqm12,zqm13,luna16}.
Therefore, longitudinal oscillations can well be described with a pendulum model. Curvature radii of magnetic dips can be exactly diagnosed,
and the lower limits of transverse magnetic field strength of the dips can be roughly estimated \citep{luna14,luna17,zqm17a,zqm17b,maz20}.
The damping mechanisms, however, are still controversial. Hydrodynamic (HD) and magnetohydrodynamic (MHD) numerical simulations have 
greatly shed light on the damping mechanisms, such as mass accretion \citep{luna12,rud16}, radiative loss \citep{zqm13}, and wave leakage \citep{zly19}.
When mass drainage takes place at the footpoints of coronal loops as a result of large initial amplitudes, the damping times are significantly reduced \citep{zqm13}.

So far, longitudinal filament oscillations triggered by flares have been substantially investigated. However, filament oscillations enhanced by flares have never been reported.
In this article, we report the multiwavelength observations of longitudinal oscillations of a filament in AR 11112 on 2010 October 18. 
The evolution of filament oscillations was divided into three phases, during which a confined C1.3 flare and an eruptive C2.6 flare associated with a jet-like CME occurred in the AR.
Data analyses are described in Sect.~\ref{s-data}. Observational results are shown in Sect.~\ref{s-res}. 
To reproduce part of the observations, we perform one-dimensional (1D) HD numerical simulations in Sect.~\ref{s-num}.
Discussions and a brief summary are arranged in Sect.~\ref{s-disc} and Sect.~\ref{s-sum}.

\section{Observations and data analysis} \label{s-data}
The filament with sinistral chirality was observed in H$\alpha$ line center by the ground-based telescope of Global Oscillation Network Group (GONG).
It was also observed in EUV wavelengths (171 and 304 {\AA}) by the Atmospheric Imaging Assembly \citep[AIA;][]{lem12} on board the Solar Dynamics Observatory (SDO).
The photospheric line-of-sight (LOS) magnetograms were provided by the Helioseismic and Magnetic Imager \citep[HMI;][]{sch12} on board SDO. 
The AIA and HMI data were calibrated using the standard Solar Software (SSW) programs \textit{aia\_prep.pro} and \textit{hmi\_prep.pro}.
The full-disk H$\alpha$ and AIA 304 {\AA} images were coaligned with a precision of $\sim$1$\farcs$2 using the cross correlation method.
Three-dimensional (3D) global magnetic fields at 12:04 UT were obtained by the potential field source surface \citep[PFSS;][]{sch03} modeling.
The 1$-$70 {\AA} flux of the flares was recorded by the Extreme Ultraviolet Variability Experiment \citep[EVE;][]{wood12} on board SDO.
Soft X-ray (SXR) fluxes of the flares in 0.5$-$4 {\AA} and 1$-$8 {\AA} were recorded by the GOES spacecraft. 
The CME associated with the C2.6 flare was observed by the C2 white light (WL) coronagraph of the
Large Angle Spectroscopic Coronagraph \citep[LASCO;][]{bru95} on board SOHO\footnote{http://cdaw.gsfc.nasa.gov/CME\_list/}.
The LASCO/C2 data were calibrated using the SSW program \textit{c2\_calibrate.pro}. The observational parameters from multiple instruments are listed in Table~\ref{table-1}.

\begin{table*}
\centering
\caption{Description of the observational parameters.}
\label{table-1}
\begin{tabular}{lcccc}
\hline\hline
Instrument & $\lambda$ &Time  & Cadence & Pixel Size \\ 
       & ({\AA})  & (UT) & (s) & (\arcsec) \\
\hline
GONG & 6562.8 & 10:55$-$22:00 & 60 & 1.0 \\
SDO/AIA  & 171, 304 & 07:00$-$22:00 &  12  & 0.6 \\
SDO/AIA  & 1600       & 07:00$-$22:00 &  24  & 0.6 \\
SDO/HMI & 6173      & 07:00$-$22:00 &  45   & 0.6 \\
SDO/EVE & 1$-$70 & 07:00$-$22:00 & 0.25 & ... \\
LASCO/C2 & WL & 16:50$-$17:48 & 720 & 11.4 \\
GOES     & 0.5$-$4 & 07:00$-$22:00 &  2.05 & ... \\
GOES     & 1$-$8    & 07:00$-$22:00 &  2.05 & ... \\
\hline
\end{tabular}
\end{table*}

\section{Observational results} \label{s-res}
\subsection{Confined flare} \label{s-flare1}
\begin{figure}
\includegraphics[width=8cm,clip=]{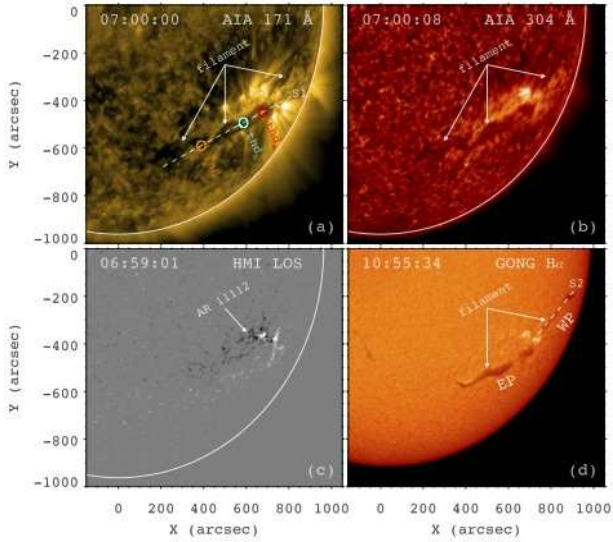}
\centering
\caption{(a)-(b) AIA 171 and 304 {\AA} images at 07:00 UT. The arrows point to the long filament. 
(c) HMI LOS magnetogram at 06:59 UT. The arrow points to AR 11112.
(d) GONG H$\alpha$ image at 10:55:34 UT. ``EP" and ``WP" represent the eastern and western parts of the filament, respectively.
In panel (a), a slice (S1) is selected to investigate the longitudinal oscillations of EP, whose time-distance diagram is drawn in Fig.~\ref{fig10}.
The red, cyan, and orange circles indicate the oscillating threads (thd$_a$, thd$_b$, and thd$_c$) in EP.
In panel (d), a slice (S2) is selected to investigate the evolution of WP, whose time-distance diagram is drawn in Fig.~\ref{fig5}.}
\label{fig1}
\end{figure}

In Fig.~\ref{fig1}, the top panels show 171 and 304 {\AA} images observed by AIA at 07:00 UT. The white arrows point to the long, dark filament residing in AR 11112.
Panel (c) shows the photospheric LOS magnetogram observed by HMI at 06:59 UT. Panel (d) shows the H$\alpha$ image observed by GONG at 10:55:34 UT.
It is clear that the filament is located along the polarity inversion line of the AR and is divided into two parts. 
The eastern part (EP) has an apparent length of $\sim$650$\arcsec$ and an angle of $\sim$26$^{\circ}$ relative to the EW direction.
The western part (WP) has an apparent length of $\sim$250$\arcsec$ and an angle of $\sim$47$^{\circ}$ relative to the EW direction.
Hence, the total length of filament reaches $\sim$900$\arcsec$. After correcting the projection effect, the true length of filament is close to the solar radius ($\sim$960$\arcsec$).

\begin{figure}
\includegraphics[width=8cm,clip=]{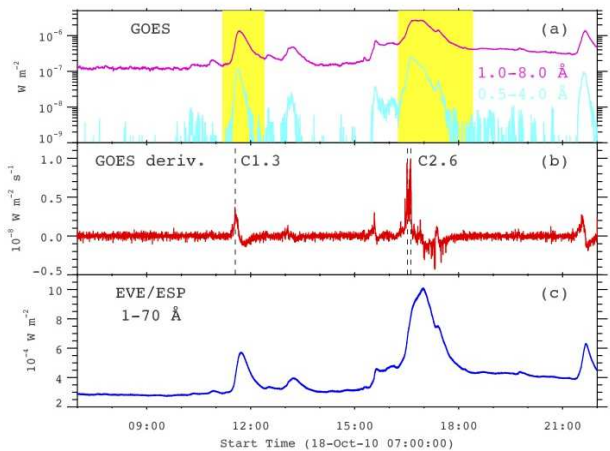}
\centering
\caption{(a) GOES SXR light curves during 07:00$-$22:00 UT on 2010 October 18.
The yellow regions denote the times of confined C1.3 flare and eruptive C2.6 flare.
(b) Time derivative of the 1$-$8 {\AA} light curve.
(c) EVE 1$-$70 {\AA} light curve.}
\label{fig2}
\end{figure}

During the whole evolution of filament, two homologous flares occurred in the AR. The first was a confined C1.3 flare. In Fig.~\ref{fig2}, SXR light curves in 0.5$-$4 and 1$-$8 {\AA} during 07:00$-$22:00 UT 
are plotted in panel (a). SXR intensities of the flare started to increase at $\sim$11:11 UT and reached the peak values at $\sim$11:39 UT before decreasing gradually until $\sim$12:24 UT.
Time derivative of the 1$-$8 {\AA} flux as a rough proxy of hard X-ray (HXR) flux is plotted in panel (b), where the first black dashed line denotes the peak time at $\sim$11:33:20 UT. 
Panel (c) shows the EVE 1$-$70 {\AA} light curve with similar evolution to the 1$-$8 {\AA} light curve except delayed peaks.

\begin{figure}
\includegraphics[width=8cm,clip=]{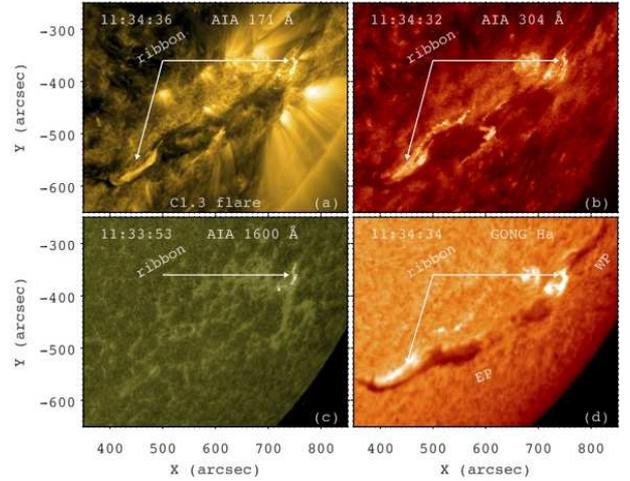}
\centering
\caption{AIA 171, 304, 1600 {\AA}, and GONG H$\alpha$ images around 11:34 UT. 
The arrows point to the bright flare ribbons. The evolution of C1.3 flare is shown in a movie (\textit{anim1.avi}) available in the online edition.}
\label{fig3}
\end{figure}

Figure~\ref{fig3} shows the AIA 171, 304, 1600 {\AA}, and GONG H$\alpha$ images around 11:34 UT.
Bright flare ribbons close to the filament are evident in all wavelengths. Since the flare was confined, both EP and WP of the filament were slightly disturbed without erupting into a CME.

\subsection{Eruptive flare and jet-like CME} \label{s-flare2}
\begin{figure}
\includegraphics[width=8cm,clip=]{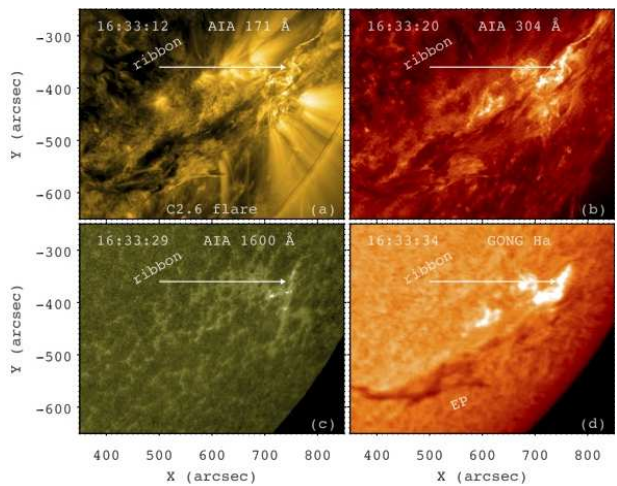}
\centering
\caption{Same as Fig.~\ref{fig3}, but for the C2.6 flare. The evolution of C2.6 flare is shown in a movie (\textit{anim2.avi}) available in the online edition.}
\label{fig4}
\end{figure}

About 5 hrs later, the C2.6 flare took place in the same AR. In Fig.~\ref{fig2}(a), SXR intensities of the flare started to increase at $\sim$16:15 UT and reached the peak values at $\sim$16:42 UT
before decreasing slowly until $\sim$18:25 UT. In Fig.~\ref{fig2}(b), two HXR peaks at 16:31:30 UT and 16:37:30 UT were associated with the flare.
Figure~\ref{fig4} shows the AIA 171, 304, 1600 {\AA}, and GONG H$\alpha$ images around 16:33 UT. The flare ribbon was cospatial with the main ribbon of C1.3 flare.
It is found that the ribbons of both flares were located near the two endpoints of the EP.

\begin{figure}
\includegraphics[width=8cm,clip=]{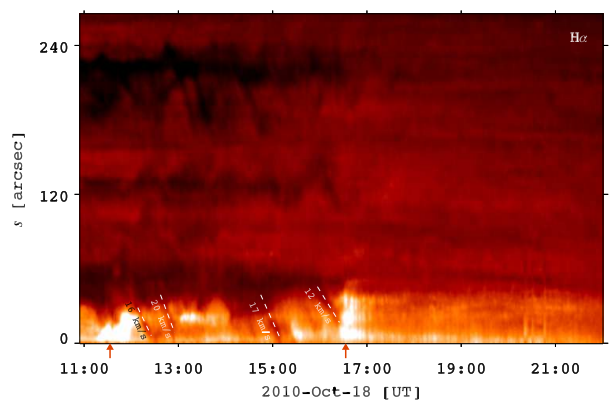}
\centering
\caption{Time-distance diagram of S2 in H$\alpha$.
$s=0\arcsec$ and $s=265\arcsec$ in $y$-axis represent the southeast and northwest endpoints of S2.
Two vertical arrows under $x$-axis indicate the times (11:33 UT and 16:33 UT) of flares.
Velocities of mass flow towards the AR are labeled.}
\label{fig5}
\end{figure}

Interestingly, the EP remained there and did not erupt during this flare, while the WP disappeared (see Fig.~\ref{fig4}(d)).
In Fig.~\ref{fig1}(d), an artificial slice (S2) along the WP with a length of 265$\arcsec$ is selected to investigate its evolution. Time-slice diagram of S2 in H$\alpha$ is displayed in Fig.~\ref{fig5}.
It is obvious that the WP was slightly disturbed during the C1.3 flare, while it was strongly disturbed during the C2.6 flare and finally erupted.
Intermittent mass flow towards the AR at speeds of 10$-$20 km s$^{-1}$ before eruption is clearly manifested in the diagram.

\begin{figure}
\includegraphics[width=8cm,clip=]{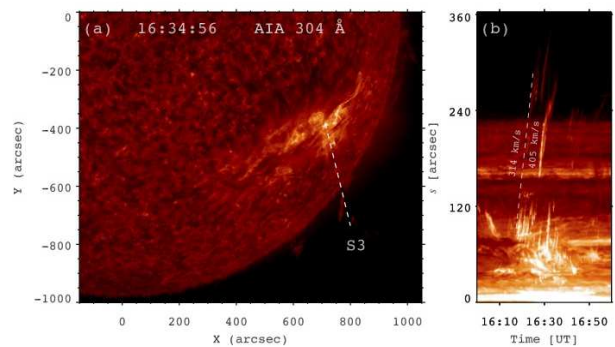}
\centering
\caption{(a) AIA 304 {\AA} image at 16:34:56 UT. 
(b) Time-distance diagram of S3 in 304 {\AA}.
$s=0\arcsec$ and $s=363\arcsec$ in $y$-axis stand for the northeast and southwest endpoints of S3.
Apparent velocities of the outflow are labeled.}
\label{fig6}
\end{figure}

Eruption of the WP in 304 {\AA} was indicated as jet-like outflow. In Fig.~\ref{fig6}, the left panel shows a snapshot of 304 {\AA} image at 16:34:56 UT.
An artificial slice (S3) with a length of 363$\arcsec$ is selected to investigate the outflow. Time-slice diagram of S3 in 304 {\AA} is displayed in the right panel.
It is seen that the outflow spurted out from $\sim$16:15 UT and lasted for about 30 minutes. The occurrence time of outflow was coincident with the SXR peak time of C2.6 flare.
The estimated apparent velocities (310$-$410 km s$^{-1}$) are comparable to those of typical coronal jets \citep{zqm14}.

\begin{figure}
\includegraphics[width=8cm,clip=]{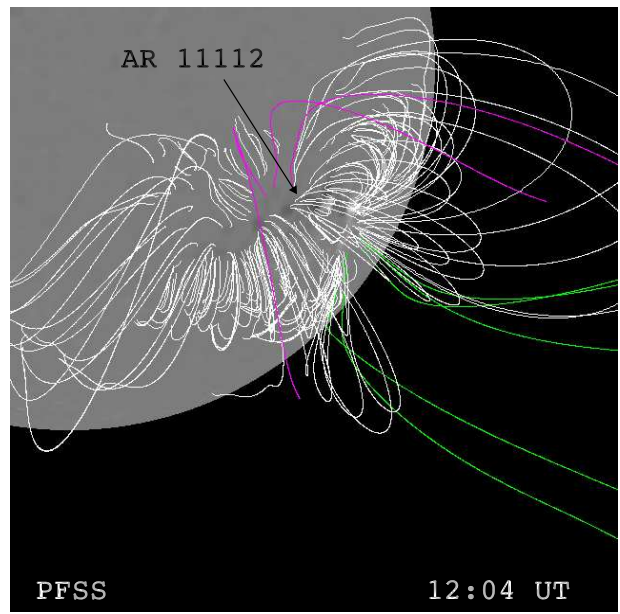}
\centering
\caption{3D magnetic field lines around AR 11112 at 12:04 UT obtained by PFSS modeling.
Open and closed field lines are coded with green/magenta and white lines.
The grayscale image denotes the LOS component of magnetic field at the photosphere.}
\label{fig7}
\end{figure}

The 3D magnetic field lines around AR 11112 at 12:04 UT are drawn in Fig.~\ref{fig7}, with open and closed field lines being coded with green/magenta and white lines, respectively.
The direction of open field (green lines) is consistent with the propagation direction of jet-like outflow, suggesting that the outflow propagated along open field lines.

\begin{figure*}
\includegraphics[width=14cm,clip=]{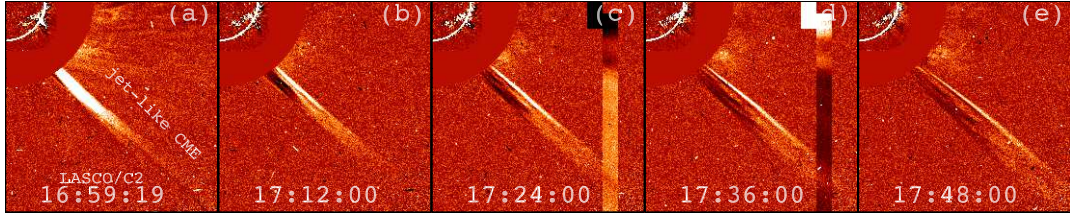}
\centering
\caption{Running-difference images of the jet-like narrow CME in the FOV of LASCO/C2.} 
\label{fig8}
\end{figure*}

\begin{figure}
\includegraphics[width=8cm,clip=]{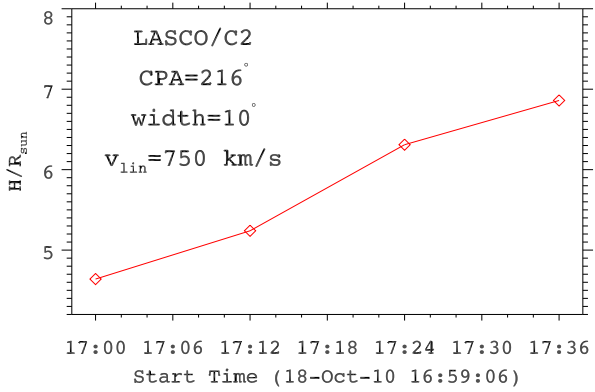}
\centering
\caption{Height-time plot of the jet-like narrow CME. $R_{sun}$ represents the solar radius.
The central position angle (CPA), angular width, and linear velocity of the CME are labeled.}
\label{fig9}
\end{figure}

WL observations of the corona from LASCO/C2 indicate that the jet-like outflow propagated even further and evolved into a narrow CME.
Figure~\ref{fig8} shows 5 snapshots of running-difference images observed by LASCO/C2.
The CME appeared at $\sim$16:59 UT and propagated until 17:48 UT in the southwest direction, which is also consistent with the direction of open field in Fig.~\ref{fig7}. 
The central position angle (CPA) and angular width are 216$^{\circ}$ and 10$^{\circ}$, respectively. 
Height-time plot of the CME is displayed in Fig.~\ref{fig9}. A linear fitting of the plot results in an apparent velocity of $\sim$750 km s$^{-1}$ in the plane of sky.
It should be emphasized that the measurements of CME heights have significant uncertainties since it is not easy to identify the leading edge.
Therefore, the derived velocity of CME is nearly twice higher than that of outflow observed in EUV wavelengths.

\subsection{Filament oscillations} \label{s-osci}
\begin{figure*}
\includegraphics[width=14cm,clip=]{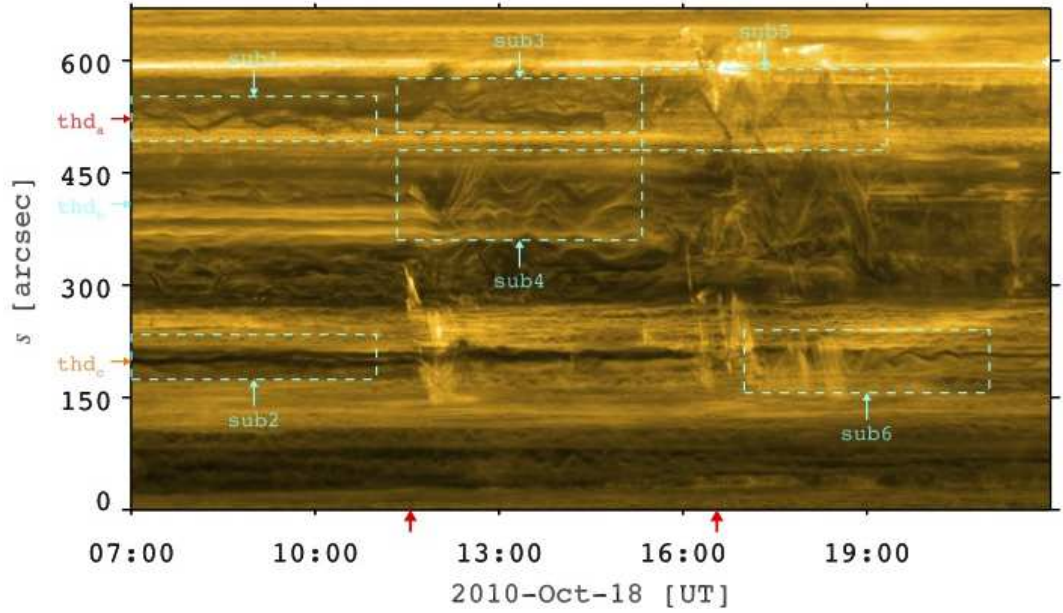}
\centering
\caption{Time-distance diagram of S1 in 171 {\AA}.
$s=0\arcsec$ and $s=669\arcsec$ in $y$-axis stand for the southeast and northwest endpoints of S1.
Horizontal arrows in red, cyan, and orange mark the positions of thd$_a$, thd$_b$, and thd$_c$, respectively.
Six subregions within the cyan boxes are extracted to explore the longitudinal oscillations of EP.
Two vertical arrows under $x$-axis indicate the times (11:33 UT and 16:33 UT) of flares.}
\label{fig10}
\end{figure*}

During 07:00$-$22:00 UT, the EP oscillated longitudinally along the filament channel. 
In Fig.~\ref{fig1}(a), at least three oscillating threads (thd$_a$, thd$_b$, and thd$_c$) could be identified and marked by red, cyan, and orange circles, respectively.
An artificial slice (S1) with a length of 669$\arcsec$ is selected to investigate the evolution of EP. Time-slice diagram of S1 in 171 {\AA} is plotted in Fig.~\ref{fig10}.
It is found that different threads oscillated in a different way, showing a very complex behavior. Six subregions (sub1-sub6) within the cyan boxes are extracted. 
Sub1 and sub2 represent oscillations before the C1.3 flare. Sub3 and sub4 represent oscillations between the two flares. Sub5 and sub6 represent oscillations after the C2.6 flare.
For thd$_a$, the oscillation was divided into three phases. Before the C1.3 flare, it oscillated with a relatively small amplitude and insignificant damping (see sub1). 
After the C1.3 flare, the amplitude increased slightly (see sub3). After the C2.6 flare, the amplitude became even larger with faster damping (see sub5).
For thd$_b$, the oscillation between the two flares was obviously demonstrated, suggesting that the oscillation was triggered by the C1.3 flare. 
After the C2.6 flare, it became chaotic and no distinct oscillation could be recognized.
For thd$_c$, the oscillations were present before the C1.3 flare and after the C2.6 flare. Between the two flares, the oscillation was not so obvious.

\begin{figure}
\includegraphics[width=8cm,clip=]{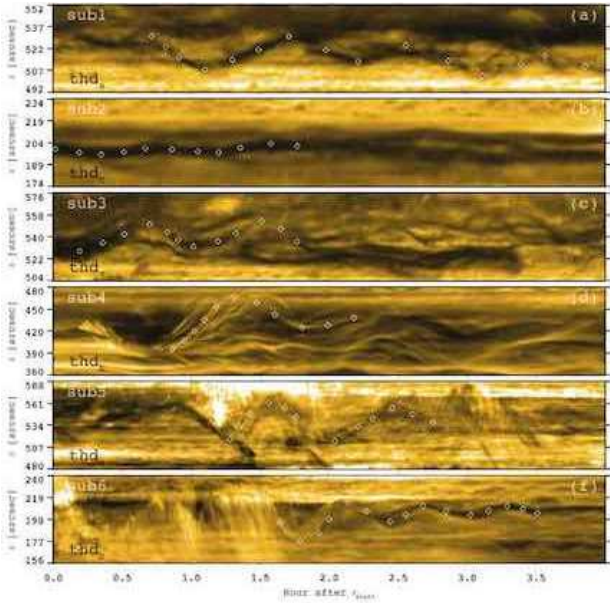}
\centering
\caption{Close-ups of the 6 subregions in Fig.~\ref{fig10}. White diamonds denote the positions of oscillating threads.
$t_{start}$ for each subregion is labeled in Table~\ref{table-2}.}
\label{fig11}
\end{figure}

\begin{figure}
\includegraphics[width=8cm,clip=]{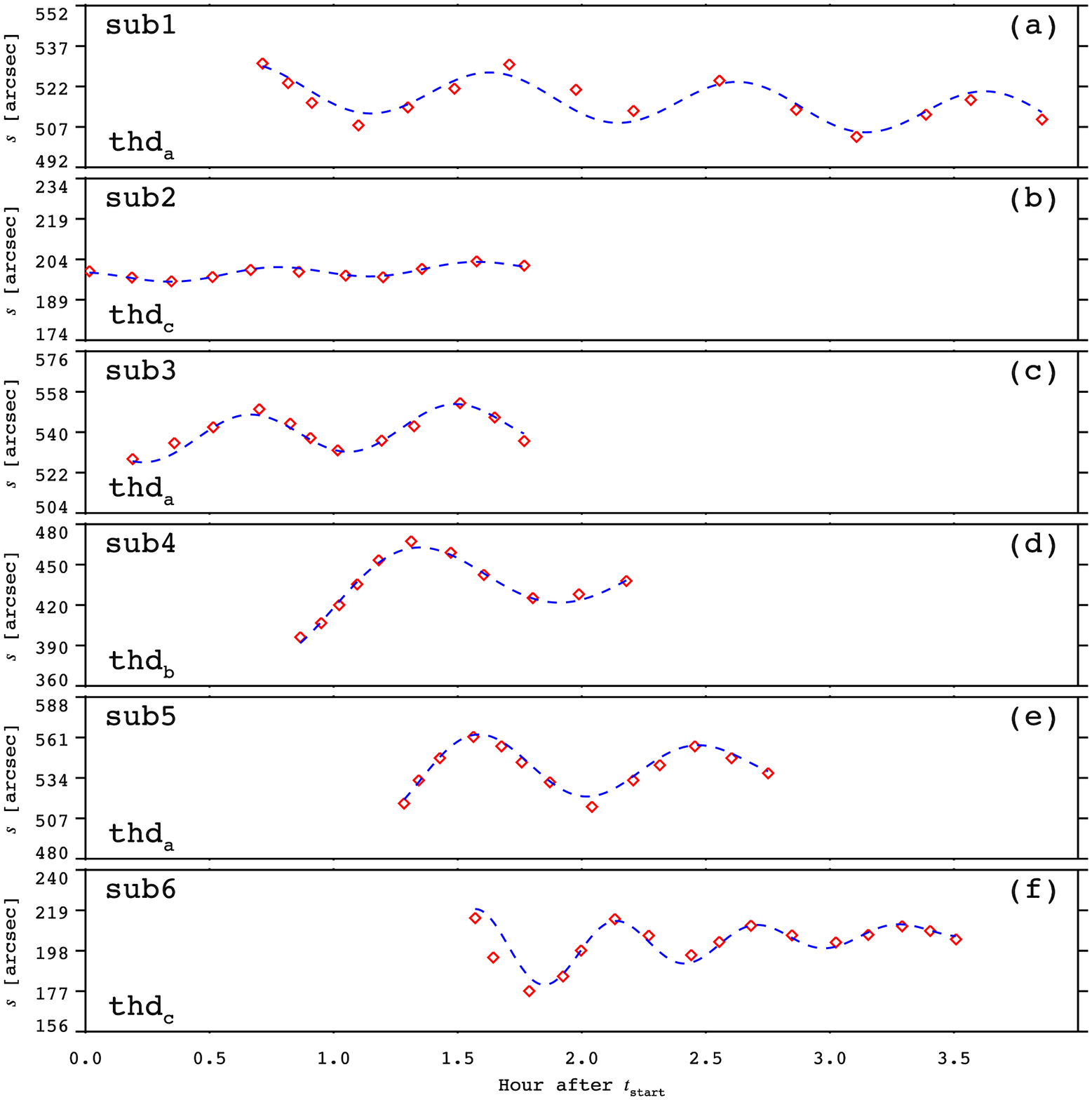}
\centering
\caption{Marked positions of the oscillating threads (red diamonds) and results of curve fitting (blue lines).
$t_{start}$ for each subregion is labeled in Table~\ref{table-2}.}
\label{fig12}
\end{figure}

In Fig.~\ref{fig11}, close-ups of the 6 subregions are shown with a better contrast. To investigate the oscillations, we track the oscillating threads and mark their positions manually with white diamonds.
The marked positions are independently plotted in Fig.~\ref{fig12}. To derive the physical parameters of oscillations, we carry out curve fittings by utilizing a sine function multiplied by an exponential term
plus a linear term:

\begin{equation} \label{eqn-1}
  y=y_0+bt+A_0\sin(\frac{2\pi}{P}t+\phi_0)e^{-t/\tau},
\end{equation}
where $y_0$, $A_0$, and $\phi_0$ represent the initial position, amplitude, and phase. $b$, $P$, and $\tau$ stand for the linear velocity, period, and damping time. The parameters are listed in Table~\ref{table-2}. 
According to the pendulum model of longitudinal filament oscillations \citep{luna12,zqm12}, the period ($P$) depends mainly on the curvature radius ($R$) of a magnetic dip, $P=2\pi \sqrt{R/g_{\odot}}$,
where $g_{\odot}=2.7\times 10^2$ m s$^{-2}$ is the gravity acceleration of the Sun. In this way, the curvature radius of an oscillating thread can be estimated, $R_{\rm Mm}=0.025 P_{\rm min}^2$, 
where the units of $R$ and $P$ are in Mm and minute, respectively. The estimated values of $R$ are listed in the last row of Table~\ref{table-2}.

From Table~\ref{table-2}, it is found that the initial amplitude ranges from 1.6 to 30 Mm with a mean value of 13.8 Mm. The amplitudes after flares are considerably larger than those before flares, 
indicating that the oscillations are greatly enhanced by flares. The initial velocity ranges from $\sim$3 to $\sim$52 km s$^{-1}$ with a mean value of $\sim$28 km s$^{-1}$.
The period ranges from 34 to 73 minutes with a mean value of $\sim$53 minutes, which is close to the period of prominence oscillation on 2007 February 8 \citep{zqm12} 
and periods of filament oscillations on 2010 August 20 \citep{luna14}. The curvature radius ranges from $\sim$29 to $\sim$133 Mm with a mean value of 73.5 Mm, 
implying that the magnetic configurations of oscillating threads differ remarkably \citep{luna14,zqm17a}. 
The oscillations with smaller amplitudes (sub1-sub3) hardly damp with time, while oscillations with larger amplitudes (sub4-sub6) damp quickly. 
The damping time ranges from $\sim$62 to $\sim$96 minutes, with a mean value of 82.4 minutes. 
$\tau/P$ has a range of 1.2$-$1.8 and a mean value of 1.6, which is comparable to those of longitudinal oscillations on 2015 June 29 \citep{zqm17b}, 
but smaller than most of the values (2$-$4) in literatures \citep[e.g.,][]{jing03,vrs07,zqm12,zqm17a,luna17}.
Observations in this study confirm our previous finding that oscillations with larger initial amplitudes tend to attenuate faster ($\tau\sim v_0^{-0.3}$) \citep{zqm13}.

\begin{table*}
\caption{Parameters of filament oscillations observed by AIA and estimated curvature radii ($R$) of the threads.}
\label{table-2}
\centering
\begin{tabular}{c|cc|cc|cc|c}
\hline\hline
  & sub1 & sub2 & sub3 & sub4 & sub5 & sub6 & average \\
  & (thd$_a$) & (thd$_c$) & (thd$_a$) & (thd$_b$) & (thd$_a$) & (thd$_c$) \\
\hline
$t_{start}$ (UT) & 07:00 & 07:00 & 11:20 & 11:20 & 15:20 & 19:20 & - \\
$t_{end}$ (UT) & 11:00 & 11:00 & 15:20 & 15:20 & 17:00 & 21:00 & - \\
$A_0$ (Mm) & 6.1 & 1.6 & 6.8 & 30.0 & 21.4 & 17.0 & 13.8 \\
$V_0$ (km s$^{-1}$) & 10.7 & 3.5 & 14.4 & 43.0 & 42.4 & 52.4 & 27.7 \\
$P$ (min) & 59.9 & 48.0 & 49.6 & 73.0 & 52.9 & 34.0 & 52.9 \\
$\tau$ (min) & - & - & - & 88.5 & 96.2 & 62.5 & 82.4 \\
$\tau/P$ & - & - & - & 1.2 & 1.8 & 1.8 & 1.6 \\
\hline
$R$ (Mm) & 89.7 & 57.6 & 61.5 & 133.2 & 70.0 & 28.9 & 73.5 \\
\hline
\end{tabular}
\end{table*}

\section{Numerical simulations} \label{s-num}
It is revealed from Fig.~\ref{fig10} that solar flares can not only trigger, but also facilitate longitudinal filament oscillations. In other words, flares can supply both energy and momentum to the filaments.
To reproduce this interesting phenomenon, especially for thd$_a$, we perform 1D HD numerical simulations following the previous works \citep{zqm12,zqm13}.
The whole evolution of a filament in a flux tube is divided into 6 steps: formation (or condensation) of cool material after chromospheric evaporation as a result of thermal instability, 
steady growth due to continuous heating and evaporation at the footpoints, relaxation into a thermal and dynamic equilibrium state after evaporation is halted, 
oscillation with a smaller amplitude after an unknown small flare, large-amplitude oscillation after flare\_1, and oscillation with a larger amplitude after flare\_2. First of all, we will briefly introduce the method.

\subsection{Simulation setup}
The 1D HD equations including optical-thin radiation are as follows \citep{xia11}: 
\begin{equation} \label{eqn-2}
  \frac{\partial \rho}{\partial t}+\frac{\partial}{\partial s}(\rho v)=0 \,,
\end{equation}

\begin{equation} \label{eqn-3}
  \frac{\partial}{\partial t}(\rho v)+\frac{\partial}{\partial s}(\rho v^2+p)=\rho g_{\parallel}(s) \,,
\end{equation}

\begin{equation} \label{eqn-4}
  \frac{\partial \varepsilon}{\partial t}+\frac{\partial}{\partial s}(\varepsilon v+pv)=\rho g_{\parallel}v+H-n_{\rm H}n_{\rm e}\Lambda(T)+\frac{\partial}{\partial s}(\kappa \frac{\partial T}{\partial s}) \,,
\end{equation}
\noindent
where $\rho$, $T$, $p$, $v$, $n_{\rm e}$, and $n_{\rm H}$ have their normal meanings ($\rho=1.4 m_{\rm p} n_{\rm H}$, $p=2.3 n_{\rm H} k_{\rm B} T$), 
$g_\parallel(s)$ is the component of gravity at a distance ($s$) along the flux tube, $\gamma=5/3$ is the ratio of specific heats, $\varepsilon=\rho v^2 /2 + p/(\gamma -1)$ is the total energy density, 
$H(s)$ is the volumetric heating rate, $\Lambda(T)$ is the radiative loss function \citep{ros78,colg08}, and $\kappa=10^{-6} T^{5/2}$ ergs cm$^{-1}$ s$^{-1}$ K$^{-1}$ is the Spitzer heat conductivity.
The above conservative equations are numerically solved using the MPI-AMRVAC\footnote{http://amrvac.org/index.html} code \citep{kepp12,por14}. 
The TVDLF scheme is adopted and a 5-level AMR is used to produce a highest resolution of 7.7 km.

\begin{figure}
\includegraphics[width=8cm,clip=]{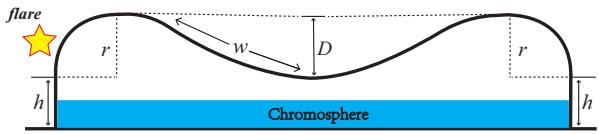}
\centering
\caption{Geometry of the flux tube used for 1D HD numerical simulations of filament oscillations.
See text for details.}
\label{fig13}
\end{figure}

Figure~\ref{fig13} shows the geometry of symmetric flux tube, consisting of two vertical legs with a length of $h$, two quarter-circular shoulders with a radius of $r$, 
and a quasi-sinusoidal-shaped dip with a length of 2$w$ and a depth of $D$ \citep{zqm12,zqm13}. Hence, the total length of the flux tube $L=2h+\pi r+2w$.
The distribution of $g_\parallel(s)$ along the tube is determined by the geometry. 
In this study, we take $h=10$ Mm, $r=10$ Mm, and $D=10$ Mm, so that the magnetic dip has a height of 10 Mm above the photosphere. 
The total length of dip is 184.68 Mm, which corresponds to a curvature radius of 90.1 Mm and an eigen period of 60.03 minutes. This is equal to the period of thd$_a$ in sub1.

\begin{figure}
\includegraphics[width=8cm,clip=]{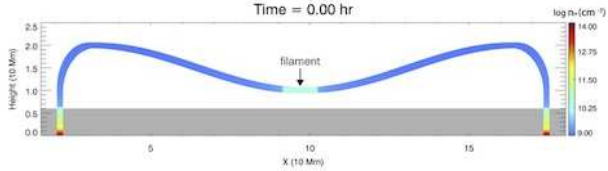}
\centering
\caption{Density distribution of the flux tube after reaching an equilibrium state. The evolution of flux tube is shown in a movie (\textit{anim3.mp4}) available in the online edition.}
\label{fig14}
\end{figure}

The first 3 steps (formation, growth, and relaxation of filament material) are the same as previous works \citep{xia11,zqm13,zhou14}. 
Figure~\ref{fig14} shows the density distribution of the flux tube after reaching an equilibrium state. The filament is located at the bottom of the dip.
The initial length, temperature, and density of thread are 12.9 Mm, 1.6$\times$10$^4$ K, and $\sim$10$^{10}$ cm$^{-3}$, respectively.
Oscillations of the thread are divided into three phases according to different impulsive heating rate $H_f(s)$:
\begin{equation} \label{eqn-5}
  H_f(s)=E_2 \exp{\left[{-\frac{(s-s_{peak})^2}{s_{scale}^2}-\frac{(t-t_{peak})^2}{t_{scale}^2}}\right]},
\end{equation}
\noindent
where the heating spatial scale $s_{scale}=2.5$ Mm, peak location $s_{peak}=15$ Mm (near the left footpoint), and heating timescale $t_{scale}=5$ minutes.
Our simulations are consistent with the observations that the energy deposition (flare ribbons) are near the footpoints of oscillating filament (see Figs.~\ref{fig3} and \ref{fig4}).
Impulsive heating, with maximal heating rate ($E_2$) of 0.015, 0.010, and 0.056 erg cm$^{-3}$ s$^{-1}$, is deposited 0.42, 3.52, and 8.52 hrs after the equilibrium state ($t=0$ hr)
to mimic a flare (unknown), a C-class flare (flare\_1), and another C-class flare (flare\_2), respectively.

\subsection{Simulation results}
\begin{figure}
\includegraphics[width=8cm,clip=]{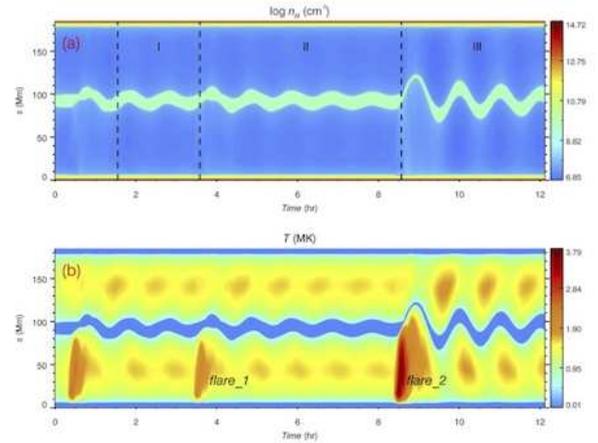}
\centering
\caption{Time evolutions of density (top panel) and temperature (bottom panel) distributions along the flux tube during filament oscillations.
$t=0$ hr signifies the start time of simulation when the flux tube reaches an equilibrium state.
Three phases of oscillations are labeled in panel (a).}
\label{fig15}
\end{figure}

\begin{figure}
\includegraphics[width=8cm,clip=]{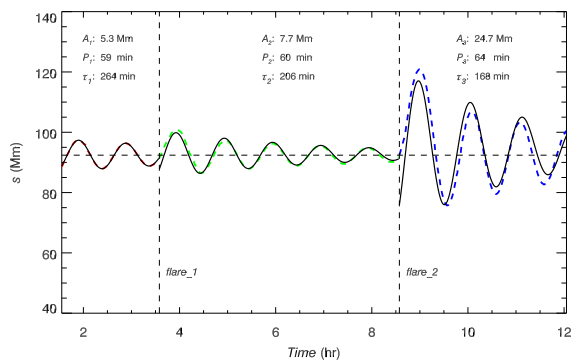}
\centering
\caption{Time evolution of filament mass center during its oscillations.
Black solid line represents the results of simulations. Red, green, and blue lines represent the results of curve fittings.
The fitted parameters, including initial amplitudes, periods, and damping times, are labeled.}
\label{fig16}
\end{figure}

In Fig.~\ref{fig15}, time evolutions of density and temperature distributions along the flux tube are plotted in the top and bottom panels.
In Fig.~\ref{fig16}, time evolution of filament mass center is plotted with black lines and the results of fitting are plotted with colored lines.
The evolution of filament oscillation is divided into three phases. In phase I, the filament oscillation is triggered by the first impulsive heating resembling an unknown flare and lasts for $\sim$3.1 hrs. 
The initial amplitude, period, and damping time are 5.3 Mm, 59 minutes, and 264 minutes, respectively (see also Table~\ref{table-3}). $\tau/P$ is close to 4.5, indicating slow attenuation.
The period is in accordance with the period of thd$_a$ in sub1, while the amplitude is slightly lower.
In phase II, the oscillation is enhanced by flare\_1 and lasts for $\sim$5 hrs. The period remains $\sim$60 minutes because the geometry of dip does not change.
The initial amplitude increases to 7.7 Mm, which is larger than that of thd$_a$ after the C1.3 flare in sub3. The damping time decreases, indicating faster attenuation ($\tau/P\approx3.4$). 
In phase III, the oscillation is enhanced by flare\_2 and lasts for 3.6 hrs. The amplitude increases significantly to 24.7 Mm, which is comparable to that of thd$_a$ after the C2.6 flare in sub5.
The period increases slightly to 64 minutes, which is predicted by the previous simulations that the period increases marginally with the initial amplitude ($P\sim v_0^{0.05}$) when the geometry is fixed \citep{zqm13}.
The damping time decreases to 168 minutes and $\tau/P$ decreases to 2.6.
It is obvious that our simulations can satisfactorily reproduce the longitudinal oscillations triggered and enhanced by solar flares. 
Both the amplitudes and periods are roughly in agreement with the results of thd$_a$.
It should be noted that the slow attenuation in phases I and II agrees with the situation of thd$_a$ before the C2.6 flare, 
while the damping time in phase III is still insufficient to explain the quick damping of thd$_a$ after the C2.6 flare. 
Since the dominant damping mechanism of longitudinal filament oscillations in our model is radiative loss \citep{zqm13}, 
a combination of radiative loss and wave leakage may be helpful in interpreting the observed quick attenuation \citep{zly19}.

\begin{table}
\caption{Parameters of simulated filament oscillations during the three phases.} 
\label{table-3}
\centering
\begin{tabular}{cccc}
\hline\hline
phase  & I & II & III \\
\hline
$\Delta t$ (hr) & 3.1 & 5.0 & 3.6 \\
$A_0$ (Mm) & 5.3 & 7.7 & 24.7 \\
$P$ (min) & 59 & 60 & 64 \\
$\tau$ (min) & 264 & 206 & 168 \\
$\tau/P$ & 4.5 & 3.4 & 2.6 \\
\hline
\end{tabular}
\end{table}

\section{Discussion} \label{s-disc}
\subsection{Relationship between flares and filament oscillations}
Since the first discovery of longitudinal filament oscillations \citep{jing03}, the triggering mechanisms have been extensively investigated, such as microflares \citep{jing06,vrs07,zqm12,zqm17a}, 
coronal jets \citep{luna14,zqm17b}, shock waves \citep{shen14b,pant16}, merging of two solar filaments \citep{luna17}, and failed filament eruption \citep{maz20}.
Longitudinal filament oscillations triggered by flares have been well established \citep{luna12,zqm13,zhou17}. However, filament oscillations enhanced by flares have never been reported.
In this paper, for the first time, we report longitudinal filament oscillations enhanced by two C-class flares in AR 11112.
The amplitudes increased from 6.1 Mm to 6.8 Mm after the C1.3 flare and further to 21.4 Mm after the C2.6 flare, with the period variation being $\leq$20\%.
The roles of flares are additionally confirmed by 1D HD numerical simulations based on the previous works. 
The simulated amplitudes and periods are close to the observed values, while the damping time in the last phase is longer than the observed value, implying additional mechanisms may play a role.
Our findings are in accordance with the daily experience of playing on a swing. Before the swing comes to a halt, the amplitude would be amplified when new momentum is deposited, 
so that the oscillation is extended dramatically.

\subsection{Relationship between filament oscillations and solar eruptions}
The relationship between large-amplitude filament oscillations and solar eruptions is still unclear. On one hand, coronal EUV waves associated with CMEs and Moreton waves associated with flares
can trigger filament oscillations \citep[e.g.,][]{eto02,dai12,liu13}.
On the other hand, after studying the transverse oscillation of a prominence using the spectroscopic observation from SOHO/SUMER,
\citet{chen08} proposed that transverse filament oscillation can serve as another precursor of CMEs, which is supported by state-of-the-art observations from IRIS \citep{zhou16}. 
After studying the longitudinal oscillation of a prominence on 2007 February 8, \citet{zqm12} proposed that longitudinal filament oscillation can serve as a new precursor of flares and CMEs, 
which is supported by the imaging observations from SDO/AIA \citep{bi14}.

In this study, the filament was divided into two parts, the EP and WP. Only the EP oscillated, lasting for about 14 hrs.
Both parts survived the C1.3 flare, while the WP erupted and produced a C2.6 flare and a jet-like CME.
Since the two parts are close to each other, it is likely that they have interplay during the evolution.
Longitudinal oscillations of the EP may stimulate the destabilization and final eruption of the WP. 
Compression from the overlying magnetic field above the EP was strong enough to prevent it from eruption after the C2.6 flare.
It should be noted that we did not perform a nonlinear force free field extrapolation, since the AR was very close to the limb.

\section{Summary} \label{s-sum}
In this paper, we investigate large-amplitude, longitudinal oscillations of the EP of a very long filament in AR 11112 observed by SDO/AIA and GONG on 2010 October 18.
HD numerical simulations using the MPI-AMRVAC code are conducted to reproduce part of the observations.
The main results are summarized as follows:
\begin{enumerate}
\item{During the evolution of filament, two homologous flares occurred in the same AR.
The C1.3 flare was confined without a CME. Both EP and WP of the filament were slightly disturbed and survived the flare.
After 5 hrs, eruption of the WP generated a C2.6 flare and a narrow, jet-like CME observed by LASCO/C2.}
\item{Three oscillating threads (thd$_a$, thd$_b$, and thd$_c$) are clearly identified in the EP and their oscillations are naturally divided into three phases by the two flares. 
The initial amplitude ranges from 1.6 to 30 Mm with a mean value of $\sim$14 Mm. The period ranges from 34 to 73 minutes with a mean value of $\sim$53 minutes. 
The curvature radii of the magnetic dips are estimated to be 29 to 133 Mm with a mean value of $\sim$74 Mm.
The damping time ranges from $\sim$62 to $\sim$96 minutes with a mean value of $\sim$82 minutes. $\tau/P$ is between 1.2 and 1.8.}
\item{For thd$_a$ in the EP, the amplitudes were enhanced by the two flares from 6.1 Mm to 6.8 Mm after the C1.3 flare and further to 21.4 Mm after the C2.6 flare.
The period variation as a result of perturbation from the flares was $\leq$20\%. The attenuation became faster after the C2.6 flare. 
To the best of our knowledge, this is the first report of large-amplitude, longitudinal filament oscillations enhanced by flares.}
\item{Numerical simulations reproduce the oscillations of thd$_a$ very well. The simulated amplitudes and periods are close to the observed values, 
while the damping time in the last phase is longer, implying additional mechanisms should be taken into account apart from radiative loss.}
\end{enumerate}

\begin{acknowledgements}
The authors appreciate the referee for valuable suggestions and comments to improve the quality of this article.
SDO is a mission of NASA\rq{}s Living With a Star Program. AIA and HMI data are courtesy of the NASA/SDO science teams. 
This work utilizes GONG data from NSO, which is operated by AURA under a cooperative agreement with NSF and with additional financial support from NOAA, NASA, and USAF.
This work is funded by NSFC grants (No. 11773079, 11790302), the Science and Technology Development Fund of Macau (275/2017/A),
the International Cooperation and Interchange Program (11961131002), the Youth Innovation Promotion Association CAS, 
and the project supported by the Specialized Research Fund for State Key Laboratories.
\end{acknowledgements}

\end{document}